\documentclass[conference]{IEEEtran}
\IEEEoverridecommandlockouts
\usepackage{amsmath,amssymb,amsfonts}
\usepackage{algorithmic}
\usepackage{textcomp}
\usepackage{xcolor}
\def\BibTeX{{\rm B\kern-.05em{\sc i\kern-.025em b}\kern-.08em
	T\kern-.1667em\lower.7ex\hbox{E}\kern-.125emX}}

\usepackage{graphicx}

\usepackage{amssymb}

\usepackage[numbers]{natbib}

\usepackage{booktabs}
\usepackage{listings}

\usepackage{csquotes}

\usepackage{graphicx}
\usepackage{epstopdf} 
\usepackage{subcaption}
\usepackage[ruled,vlined,linesnumbered]{algorithm2e}

\SetCommentSty{mycommfont}

\SetAlFnt{\small}
\SetAlCapFnt{\small}
\SetAlCapNameFnt{\small}
\SetAlCapHSkip{0pt}
\IncMargin{-\parindent}

\lstdefinestyle{CPP}{
frame=single,  breaklines=true, basicstyle=\scriptsize,
numbers=left, numberstyle=\tiny, stepnumber=1, numbersep=5pt,%
backgroundcolor=\color{gray!10},%
}%

\usepackage{xcolor}
\usepackage{listings}

\lstdefinestyle{ANTLR}{
numbers=left, numberstyle=\tiny, stepnumber=1, numbersep=5pt, frame=single,
backgroundcolor=\color{gray!10},%
basicstyle=\small\ttfamily\color{black},%
breaklines=true,
moredelim=[s][\color{green!50!black}\ttfamily]{'}{'},
moredelim=*[s][\color{black}\ttfamily]{options}{\}},
commentstyle={\color{gray}\itshape},
morecomment=[l]{//},
emph={%
	STRING
},emphstyle={\color{blue}\ttfamily},
alsoletter={:,|,;},%
morekeywords={:,|,;},
keywordstyle=\bfseries\color{black},
moredelim=[is][keywordstyle]{|>}{<|},%
}

\begin{document}

\title{HSTREAM: A directive-based language extension for heterogeneous stream computing \thanks{PREPRINT, CSE 2018, \copyright IEEE}}

\author{\IEEEauthorblockN{Suejb Memeti}
	\IEEEauthorblockA{\textit{Department of Computer Science} \\
		\textit{Linnaeus University}\\
		V\"{a}xj\"{o}, Sweden \\
		suejb.memeti@lnu.se}
	\and
	\IEEEauthorblockN{Sabri Pllana}
	\IEEEauthorblockA{\textit{Department of Computer Science} \\
		\textit{Linnaeus University}\\
		V\"{a}xj\"{o}, Sweden \\
		sabri.pllana@lnu.se}
}

\date{Received: date / Accepted: date}

\maketitle

\begin{abstract}
	Big data streaming applications require utilization of heterogeneous parallel computing systems, which may comprise multiple multi-core CPUs and many-core accelerating devices such as NVIDIA GPUs and Intel Xeon Phis. Programming such systems require advanced knowledge of several hardware architectures and device-specific programming models, including OpenMP and CUDA.
	In this paper, we present HSTREAM, a compiler directive-based language extension to support programming stream computing applications for heterogeneous parallel computing systems. HSTREAM source-to-source compiler aims to increase the programming productivity by enabling programmers to annotate the parallel regions for heterogeneous execution and generate target specific code. The HSTREAM runtime automatically distributes the workload across CPUs and accelerating devices.
	We demonstrate the usefulness of HSTREAM language extension with various applications from the STREAM benchmark.
	Experimental evaluation results show that HSTREAM can keep the same programming simplicity as OpenMP, and the generated code can deliver performance beyond what CPUs-only and GPUs-only executions can deliver.
\end{abstract}


\section{Introduction}
\label{sec:introduction}

Nowadays, a huge amount of data is generated throughout various mechanisms, such as scientific measurement and experiments (including genetics, physics, and astronomy), social media (including Facebook, and Twitter), and health-care \cite{dobre2014,memeti2015analyzing}. The current challenges of big data include storing and processing very large files.

However, in big data applications, not necessarily the entire data has to be processed at once. Furthermore, in most of the big-data applications, the data may be streamed, which means flowing in real-time (for instance data coming from different sensors in the Internet of Things), and therefore, it may not be available entirely. In such cases, the data needs to be processed in chunks and continuously \cite{mizell2017}.

Heterogeneous parallel computing systems comprise multiple non-identical processing units (PU), including CPUs on the host and accelerating devices (such as GPU, Intel Xeon Phi, and FPGA). Most of the top supercomputers in the world \cite{top500} comprise multiple nodes with heterogeneous processing units. For instance, the nodes of the current number one supercomputer in the TOP500 list consist of two IBM POWER9 CPUs and six NVIDIA Volta V100 GPUs.

While the combination of such heterogeneous processing units may deliver high performance, scalability, and energy efficiency, programming and optimizing such systems is much more complex \cite{benkner2011peppher,Memeti2018,Czarnul2017}. Different manufacturers of accelerating devices prefer to use different programming frameworks for offloading (which means transferring the data and control from the host to the device). For instance, OpenMP is used to offload computations to Intel Xeon Phi accelerators, whereas CUDA and OpenCL are used for offloading computations to GPUs.

The programming complexity of such systems leads to system underutilization. For example, applications designed for multi-core processing are able to utilize the available resources on the host CPUs. However, while the host CPUs are performing the actual work, the accelerating devices remain idle. On the other hand, most of the applications designed for accelerating devices use the CPU resources just for performing the data transfers and initiation of kernels, which is often performed by a single thread. Modern multi-core CPUs comprise a larger number of cores/threads, hence most of the CPU resources remain idle.

Researchers have proposed different techniques to address challenges of heterogeneous parallel programming and big data. For instance, source-to-source compilation techniques are proposed to ease programming of data-parallel applications \cite{yan2017homp,fonseca2013,ansel2009petabricks,rossbach2013}. Similarly, approaches that are based on C++ template library are used for stream and data parallel computing systems \cite{del2017generic,ernstsson2018}. \citet{pop2011stream} propose a language extension to OpenMP for stream computing on multi-core architectures. To the best of our knowledge, there does not exist any source-to-source compiler that supports stream computing for heterogeneous parallel computing systems.

\begin{figure}[ht]
	\centering
	\includegraphics[width=\linewidth]{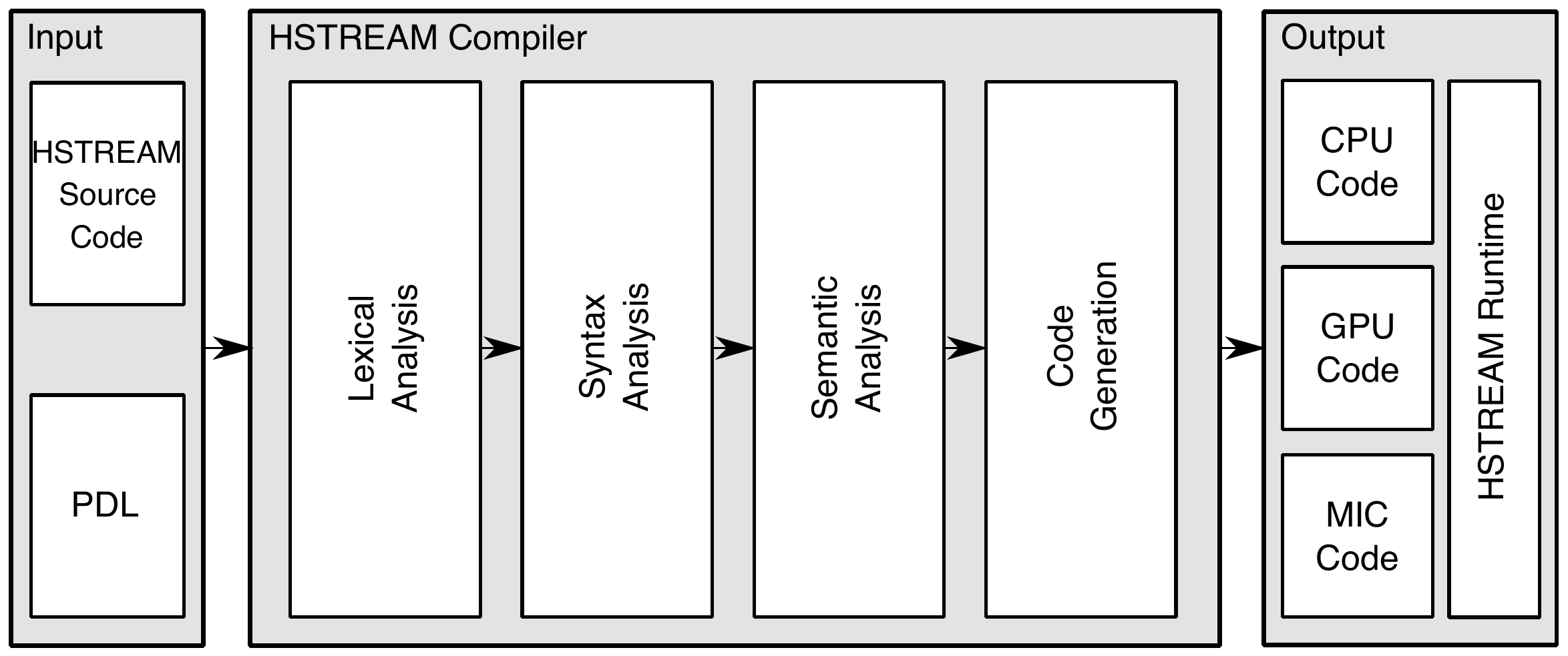}
	\caption{Overview of the proposed approach.}
	\label{fig:overview}
\end{figure}

In this paper, we present HSTREAM, a compiler directive-based language extension that supports heterogeneous stream computing. HSTREAM aims to keep the same simplicity as programming with OpenMP and to enable programmers to easily utilize the available heterogeneous parallel computing resources on the host (CPU threads) and device (GPUs, or Intel Xeon Phis). The overview of the HSTREAM solution is depicted in Fig. \ref{fig:overview}. The HSTREAM source-to-source compiler performs several analysis steps (including lexical, syntactical, and semantical) and generates target specific code from a given source code annotated with HSTREAM compiler directives and a PDL file that describes the hardware architecture. HSTREAM supports code generation for multi-core CPUs using OpenMP, GPUs using CUDA, and Intel Xeon Phis (also known as MIC) using Intel Language Extensions for Offloading (LEO). The HSTREAM runtime is responsible for scheduling the workload across the heterogeneous PUs.

We use the HSTREAM source-to-source compiler to generate the heterogeneous version of the STREAM benchmark \cite{McCalpin1995}. We evaluate the generated heterogeneous STREAM benchmark with respect to programming productivity and performance. The experimental results show that HSTREAM keeps the same simplicity as OpenMP, and the code generated for execution on heterogeneous systems delivers higher performance compared to CPUs-only and GPUs-only execution.

Major contributions of this paper include:

\begin{itemize}
	\item HSTREAM compiler - a source-to-source compiler for generating target specific code from high-level directive-based annotated source code.
	\item HSTREAM runtime - a runtime system for scheduling the workload across various non-identical processing units.
	\item Evaluation of the usefulness of HSTREAM using applications from the STREAM and STREAM2 benchmarks.
\end{itemize}

The rest of the paper is structured as follows. The design, execution model, and implementation aspects of HSTREAM are described in Section \ref{sec:hstream}. Section \ref{sec:evaluation} describes the experimental environment (including the system configuration, STREAM benchmark, and evaluation metrics) and experimental results. We compare and contrast our work with the current state-of-the-art in Section \ref{sec:rw}. Section \ref{sec:conclusion_fw} concludes this paper and provides information about future work.

\section{HSTREAM: Language extension to support heterogeneous stream computing}
\label{sec:hstream}

In this section, we describe the design, the execution model, and the implementation aspects of HSTREAM.

\subsection{Design}
\label{sec:hstream_design}

OpenMP 4.5 supports offloading of computations to accelerators. However, a single loop is usually offloaded to a single device, and while one device is performing some computations, the other PUs (including host CPUs and accelerators) remain idle. Distributing the data and computations of the same loop across multiple accelerating devices and host CPUs requires additional programming investment.

HSTREAM enables the automatic distribution of data and computations across different PUs. It enables programmers to easily exploit the available resources in heterogeneous parallel computing systems. The source-to-source code generation helps to reduce the programming time investment and errors that may come when explicitly handling communication and synchronization between various processing units.

While HSTREAM is designed for heterogeneous computing of stream applications, it can also be used for data-parallel applications. In the context of HSTREAM, data-parallel applications process data that is entirely available in memory, whereas stream computing process some data that is streamed from an I/O device. Streams are read in chunks, stored in local buffers first, processed by multiple heterogeneous PUs, and transformed back to a stream or stored somewhere in memory.

Listing \ref{lst:hstream_directive} shows the syntax of the HSTREAM compiler directive, which starts with the \textit{\#pragma} keyword followed by the \textit{hstream} keyword that stands for heterogeneous stream. Thereafter, multiple clauses can occur in any order. Details about each of the directive clauses are provided below.

\begin{lstlisting}[style=CPP,escapechar=|, caption={An example of the HSTREAM compiler directive.}, label={lst:hstream_directive}]
#pragma hstream in(...) out(...) inout(...) device(...) scheduling(...) |\label{lst:hstream_directive_pragma}|
{
	//body
}
\end{lstlisting}

\textbf{\textit{In} clause:} The \textit{in clause} is used to indicate the data that should be transferred to the accelerating devices for processing. The syntax for the \textit{in clause} is inspired from the Intel LEO and looks as follows: $in(variable\_ref~ [,~variable\_ref~...])$. The \textit{variable\_ref} can be a simple data type, array, or stream. For example, $in(a)$ where $a$ may be either a simple data type or an array, and $in(a:double)$ where $a$ is a stream of data of type double. The in clause may accept multiple variables of different types. For instance, in $in(a,~b,~c:int)$, the first variable ($a$) is an array, $b$ is a scalar value, and $c$ is a stream of integers.
The \textit{in clause} can be used multiple times within the same directive. For instance, $in(a,~b)~in(c:int)$ is equivalent to the example above.

\textbf{\textit{Out} clause:} The out clause is used to indicate the variables that need to be transferred from the accelerators to the host memory. The syntax for the \textit{out clause} is similar to the \textit{in clause} and looks as follows: $out(variable\_ref~ [,~variable\_ref~...])$.

\textbf{\textit{InOut} clause:} The \textit{inout clause} is used to indicate the variables that need to be transferred to and back from the accelerating devices. The syntax looks as follows: $inout(variable\_ref~ [,~variable\_ref~...])$. The \textit{inout clause} combines the functionality of the \textit{in} and \textit{out clause}. For example, $inout(a)$ has the same functionality as using the \textit{in clause} and the \textit{out clause} separately ($in(a)~out(a)$).

\textbf{\textit{Device} clause:} In comparison to the existing OpenMP \textit{device clause}, which is used to specify only one accelerating device id as offloading target, the \textit{device clause} of the HSTREAM language extension allows providing a list of PU ids that will collaboratively process the input data elements provided using the \textit{in, out,} and \textit{inout} clauses. The syntax of the \textit{device clause} looks as follows: $device(device\_id~ [,~device\_id~...])$. The following examples describe different scenarios of the use of \textit{device clause}: (1) $device(*)$ is the default value of the \textit{device clause}, which means that all PUs should be used including the CPUs and accelerating devices; (2) $device(1)$ means that only PU with id 1 should be used; (3) $device(0,1,2)$ means that PUs with id 0, 1 and 2 will work together to process some data.
Details for each PU (such as, id, number of cores, cache size, core frequency, and global memory) are extracted from the platform description file (.pdl), which needs to be provided as input to our compiler. The PDL is a platform description language for the explicit description of heterogeneous parallel computing systems \cite{sandrieser2011pdl}.

\textbf{\textit{Scheduling} clause:} The \textit{scheduling clause} is used to determine the sizes of data chunks for each PU. There are different scenarios on how to use the \textit{scheduling} clause: (1) Device specific distribution, where the programmer will explicitly set the chunk size for each processing unit. The syntax looks as follows, $scheduling(device\_id:chunk\_size~[,~device\_id:chunk\_size~...])$; (2) Uniform distribution, where the programmer explicitly sets a constant uniform distribution for all PUs. The syntax looks as follows, $scheduling(chunk\_size)$; and (3) Automatic distribution, where the HSTREAM runtime system automatically determines the chunk sizes for each of the PU. The syntax for automatic distribution looks as follows, $scheduling(AUTO)$.

\subsection{The execution model}
\label{sec:execution_model}

\begin{figure}[ht]
	\centering
	\includegraphics[width=\linewidth]{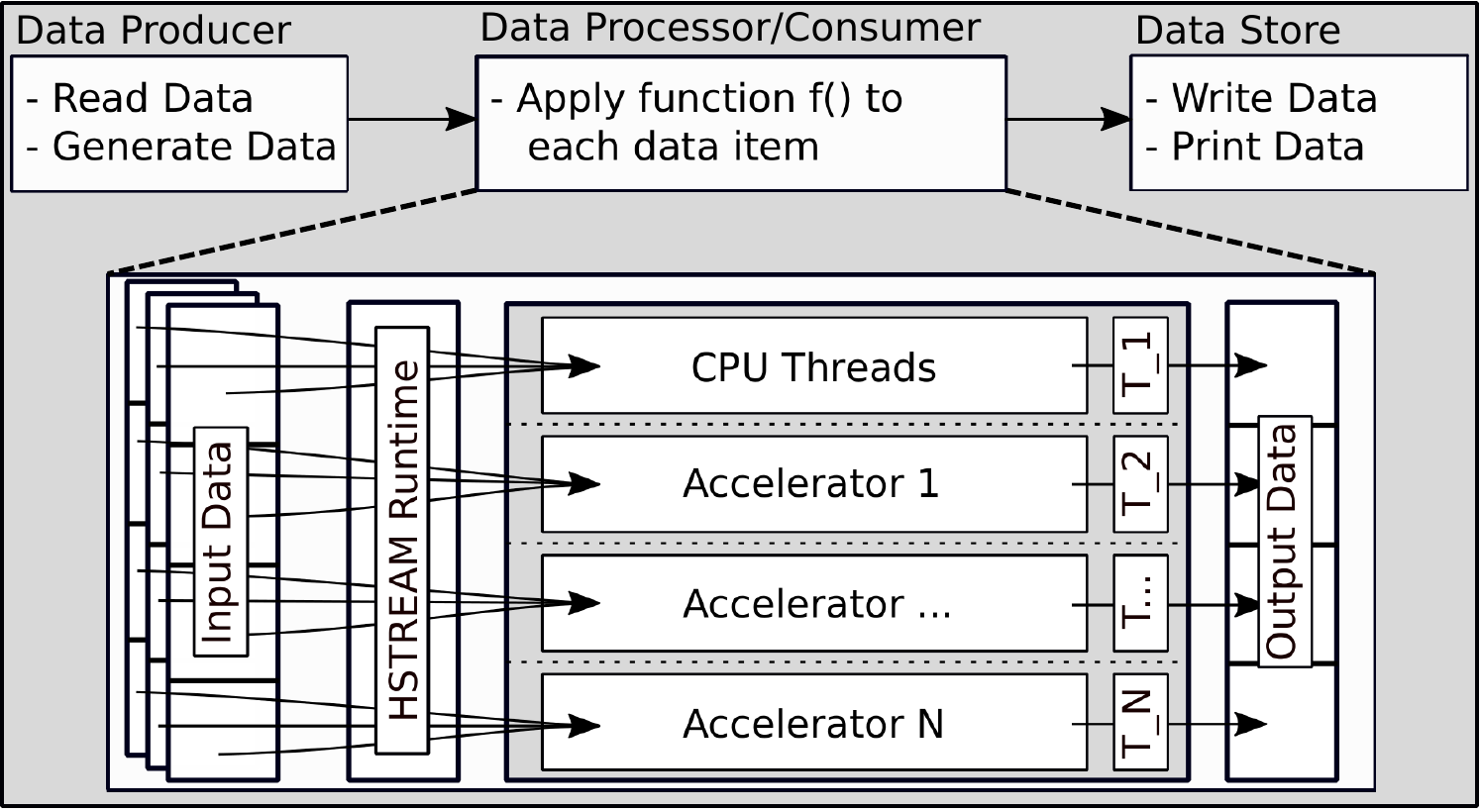}
	\caption{The execution model of HSTREAM solution for stream computing in heterogeneous parallel computing systems.}
	\label{fig:execution_model}
\end{figure}

Figure \ref{fig:execution_model} depicts an overview of the HSTREAM execution model. The main components of our solution are, the data producer, data processor, and data store. The data producer reads the data in batches from an input stream (that can be a file, data coming from the network, ...). The data processor stores the batches locally and then applies a specified function to each data item. Once the data is processed (consumed), the output is sent to the third component (named data store) to either write the data to a file or print it.

\begin{figure}[ht]
	\centering
	\includegraphics[width=\linewidth]{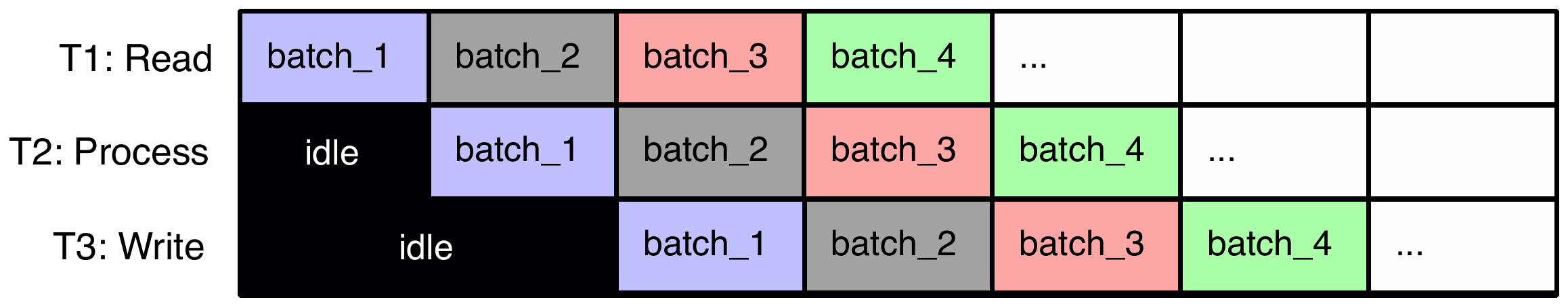}
	\caption{An example of overlapping data read, data process, and data write.}
	\label{fig:data_read_process_write_overlap}
\end{figure}

Please note that this process can be overlapped, which means that while data producer is reading one batch, the data consumer can process another batch, and at the same time the data store can write another batch to a file. Figure \ref{fig:data_read_process_write_overlap} shows an example of overlapping the reading, processing, and writing of data. To achieve this result, we need three separate threads, T1 is responsible to read the data, T2 is responsible to initiate the data processing, and T3 writes the data. First, the data producer starts reading the first batch and then notifies T2 that there is data available for processing. While T2 reads data, T1 can continue reading the next batch. When T2 finishes processing the first batch, it will notify T3 that there is data ready to be written. Then, T3 will start writing the output data to a file. While T3 is writing the first batch, T2 may start processing the second batch (only if T1 has finished reading batch 2), and T1 can start reading batch 3.

The data processor in Fig. \ref{fig:execution_model} uses all the available PUs (or specific ones, depending on what the developer has provided in the scheduling clause) to process the input data. There is a separate CPU thread controlling each PU, including (remaining) CPU threads, GPUs, and Intel Xeon Phis. Each CPU thread (except the one that controls the CPU threads) is responsible to transfer the corresponding data chunk from the host local storage to the memory of the accelerating device. The data will be processed in the accelerating devices, and then transferred back to the host. The thread that controls the remaining CPU threads does not need to do any explicit data transfer, because the controlling thread and the processing threads share the same DRAM. Please note that the same overlapping strategy used for reading, processing, and writing data can be used within the data processing component, such that the data transfer (host-to-device, and device-to-host) is overlapped with the data processing.

\subsection{Implementation}
\label{sec:hstream_implementation}

In this section, we will describe the tools and techniques used to implement our source-to-source compiler. Thereafter, throughout an example, we will describe the transformation process from a high-level C++ code annotated with HSTREAM compiler directives, to C++ code with OpenMP directives for execution on host CPUs, Intel LEO for execution on Intel Xeon Phi coprocessors, and CUDA for execution on GPU accelerators.

\subsubsection{Source-to-Source Compiler}

\begin{figure}[t]
	\centering
	\includegraphics[width=\linewidth]{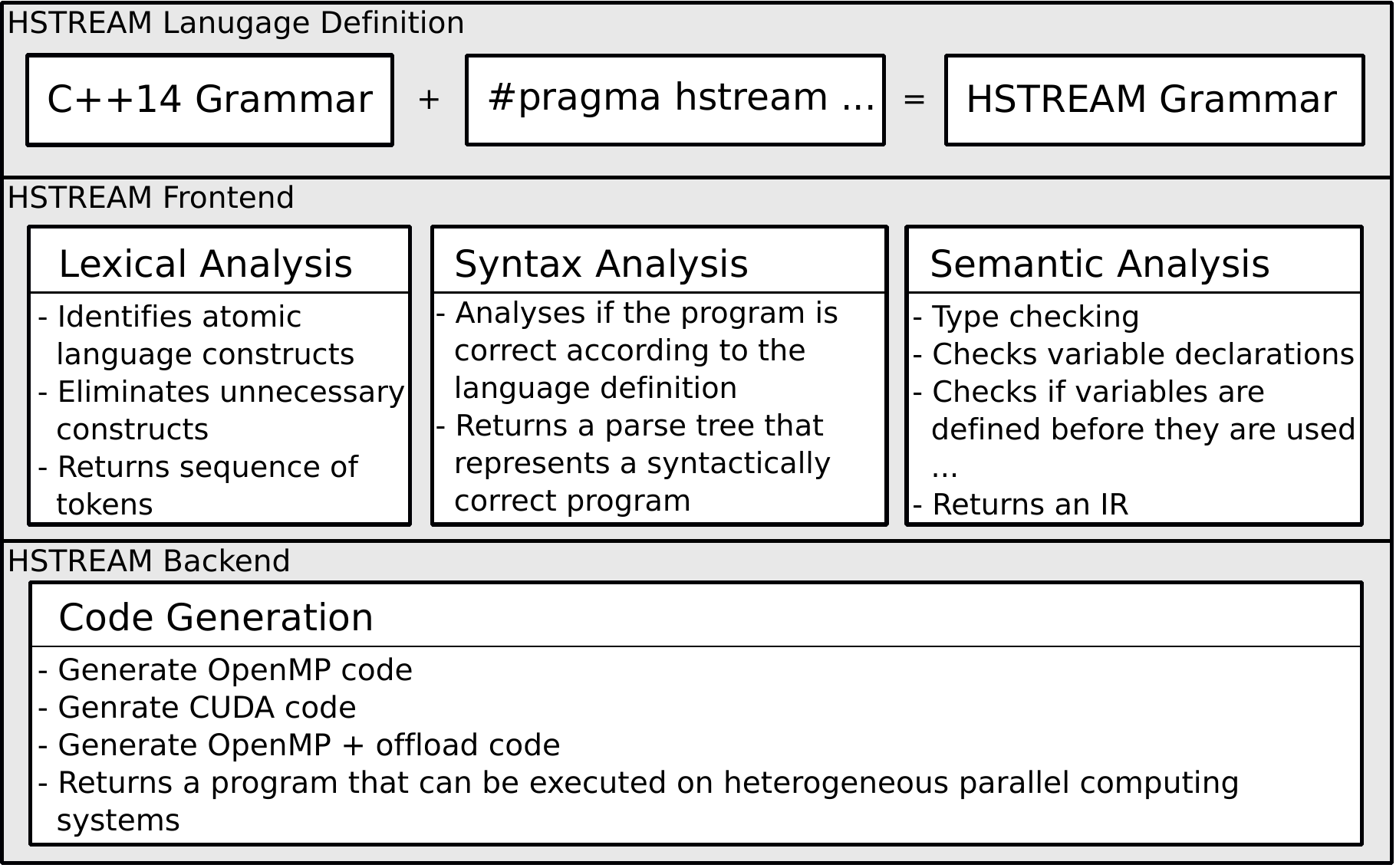}
	\caption{Overview of the implementation of our source-to-source compiler.}
	\label{fig:compiler_overview}
\end{figure}

Figure \ref{fig:compiler_overview} depicts an overview of the implementation steps, including the HSTREAM language definition, front-end, and back-end.

We use ANTLR4 (Another Tool for Language Recognition) \cite{parr2013definitive} to write the grammar for our compiler, that we call \textit{HSTREAM grammar}. As a basis for our work, we used the C++14 grammar from ANTLR Project GitHub repository of a collection of ANTLR4 grammars \cite{antlrgrammars}. We extend the C++14 grammar with the \textit{HSTREAM} compiler directives to allow developers to annotate parts of the code that need to be executed in heterogeneous parallel computing systems. Listing \ref{lst:hstream_grammar} shows an excerpt of the HSTREAM grammar written in ANTLR4.

\begin{lstlisting}[style=ANTLR,caption={An excerpt of the HSTREAM Grammar.},label={lst:hstream_grammar}]
|>directives :<| directive* ;
|>directive  :<| PRAGMA HSTREAM clauses body ;
|>clauses    :<| clause* ;
|>clause	   :<| inclause
	   | outclause
	   | inoutclause
	   | deviceclause
	   | schedulingclause
	   ;
\end{lstlisting}

The ANTLR tool is used to generate the parser, which is used for lexical, syntax, and semantic analysis. The lexical analyzer takes the input file, identifies atomic language constructs (with the help of regular expressions and pattern rules), removes any unnecessary constructs, such as comments and white spaces, and returns a sequence of tokens. In the case of illegal language constructs (tokens) errors will be generated.

The sequence of tokens is passed to the syntax analyzer (also known as the parser) to check (with the help of context-free grammars) if the input source code is correct according to the language definition. The parser generates a parse tree, which represents a syntactically correct program.

The generated parse tree is used by the semantic analyzer to judge whether the syntax derives any meaning. Some of the main tasks associated with the semantic analysis include type checking, scope resolution, checking for variable declarations, and checking whether variables are defined before they are used. If there are semantic errors, appropriate feedback will be provided, such as type mismatch, undeclared variables, multiple variable declarations within a scope, and variable access is out of scope. For instance, an HSTREAM directive with two or more \textit{scheduling} or \textit{device} clauses is syntactically correct, but semantically wrong, because only one such clause can be provided. If no semantic errors are identified, then an intermediate representation (IR) of the source code will be created, which may be used for optimization and translation.

We do not perform any optimization of the code during source-to-source compilation, but we use the IR to generate the target-specific code. We use the String Template (ST) Library \cite{parr2004enforcing} to generate the target source code. A simple example of the ST library that is used to generate the CUDA memory transfer statements is shown in Listing \ref{lst:string_template}.

\begin{lstlisting}[style=ANTLR,numbers=none,caption={An example of the string template library that is used to generate the CUDA memory transfer statements.},label={lst:string_template}]
cuda_memcpy_host_to_device(from, to, type) ::= "cudaCheckError(cudaMemcpy($from$, d_$to$, sizeof($type$)*myN, cudaMemcpyHostToDevice));"
\end{lstlisting}

Based on the type of the system architecture and based on the type of PUs provided in the \textit{device clause} the corresponding functions will be generated. At the time of writing this paper, HSTREAM source-to-source compiler supports code generation for OpenMP, CUDA, and Intel LEO.

\subsubsection{Code transformation example}

In this section, throughout an example we describe the transformation process from high-level C++ code annotated with HSTREAM directives, to target specific source code. The source code for the \textit{TRIAD} function written in C++ with HSTREAM annotations is shown in Listing \ref{lst:hstream_triad}. Line \ref{lst:hstream_triad_pragma} shows the HSTREAM pragma directive, which includes the \textit{hstream} keyword, the \textit{in} and \textit{out clause} for memory management, and the \textit{device} and \textit{scheduling clause} for scheduling of the workload. The \textit{TRIAD} function takes three array variables and one scalar variable as input and outputs the result to one of the arrays. In this example, we want to use all available PUs (\textit{device(*)}), and we want a uniform distribution of the workload for each of the PUs (\textit{scheduling(4096)}).

\begin{lstlisting}[style=CPP,escapechar=|, caption={HSTREAM TRIAD function of the STREAM benchmark.}, label={lst:hstream_triad}]
#pragma hstream in(b,c,a,scalar) out(a) device(*) scheduling(4096) |\label{lst:hstream_triad_pragma}|
{
    a = b+scalar*c;
}
\end{lstlisting}

We assume that we have a heterogeneous platform that comprises CPUs, GPUs, and Intel Xeon Phi co-processor. Please note that we describe our hardware architecture using XML-based platform description language \cite{sandrieser2011pdl}. The HSTREAM compiler will generate three types of functions, each designed for execution on one of the PUs. For example, an OpenMP based function will be generated for execution on multi-core CPUs, a CUDA kernel will be generated for execution on GPU devices, and another pragma-based function that uses Intel LEO for execution on Intel Xeon Phi.

Listing \ref{lst:CPU_triad} shows the TRIAD code generated for execution on host CPUs.

\begin{lstlisting}[style=CPP, caption={TRIAD function of the STREAM benchmark designed for execution on multi-core CPUs.}, label={lst:CPU_triad}]
#pragma omp parallel for
for (int i=start; i<finish; i++)
{
    a[i] = b[i]+scalar*c[i];
}
\end{lstlisting}

The corresponding function for execution on the GPU accelerators is shown in Listing \ref{lst:GPU_triad}.
Please note that Listing \ref{lst:GPU_triad} shows only the CUDA kernel, whereas the code for memory management, such as allocation, the transfer from host to device and vice-versa, and the memory deallocation is handled by our runtime scheduler (shown in Algorithm \ref{alg:scheduler}).

\begin{lstlisting}[style=CPP, caption={TRIAD function of the STREAM benchmark designed for execution on GPUs.}, label={lst:GPU_triad}]
__global__ void GPU_Triad( double *b, double *c, double *a, double scalar, int len) {
    int idx = threadIdx.x + blockIdx.x * blockDim.x;
    if (idx < len)
    {
        a[idx] = b[idx]+scalar*c[idx];
    }
}
\end{lstlisting}

Listing \ref{lst:MIC_triad} shows the generated source code for execution on the Intel Xeon Phi (also known as MIC), which corresponds to the \textit{TRIAD} function from the STREAM benchmark.

\begin{lstlisting}[style=CPP, escapechar=|, caption={Triad function of the STREAM benchmark designed for execution on Intel Xeon Phi.}, label={lst:MIC_triad}]
#pragma offload target(mic: cpu_thread_id) in(a[my_start:my_finish]) in(c[my_start:my_finish])  out(c[my_start:my_finish]) |\label{lst:MIC_triad_directive}|
{
    #pragma omp parallel for
    for (int  i = my_start; i < my_finish; i++)
    {
        a[i] = b[i]+scalar*c[i];
    }
}
\end{lstlisting}

The pseudo-code for the runtime scheduler, which is responsible for distributing the workload across the heterogeneous processing units, is shown in Algorithm \ref{alg:scheduler}. The generated runtime class has information for the hardware architecture, which is derived from the provided platform description file (.pdl).
This class, together with the information provided in the \textit{device clause}, are used to determine how many threads we need to engage. Since one thread controls a separate PU, we create as many threads as there are PUs. In the initialization  step (see Line \ref{alg:scheduler_initialization}) we create an instance to the runtime class that has information about the system and create two shared variables that keep track of the current state (start and finish positions) of the processed data.

Each CPU thread controls a PU, and each thread is responsible to determine the \textit{start} and \textit{finish} index of its own chunk of data (see Line \ref{alg:scheduler_chunk_determination}). To avoid multiple PUs processing the same amount data, we need to perform this process in a critical section, which means that while one thread is determining its \textit{start} and \textit{finish} positions, no other thread can do the same. Furthermore, thread private variables of the \textit{start} and \textit{finish} variables are created, to enable other threads to pick other chunks of data while another one is processing its chunk of data.

Once the \textit{start} and \textit{finish} positions are determined, the data is ready for processing. Using a single \textit{if-then-else} statement we check the type of the PU (see Line \ref{alg:scheduler_pu_type_GPU}, \ref{alg:scheduler_pu_type_MIC}, and \ref{alg:scheduler_pu_type_CPU}) and perform the corresponding steps for each type. For instance, for GPU accelerated devices, we need explicit data management, such as device memory allocation, transferring data from host to device and vice-versa, and memory deallocation (see Line \ref{alg:scheduler_create_device_variables}-\ref{alg:scheduler_gpu_finish}). Similarly, for Intel Xeon Phi accelerators, explicit data management is performed through the \textit{in} and \textit{out} clauses of the Intel LEO directive (see Listing \ref{lst:MIC_triad} Line \ref{lst:MIC_triad_directive}). For CPUs, there is no need for explicit data transfer because the controlling thread and the processing threads share the same DRAM (see Listing \ref{lst:CPU_triad}).

\begin{algorithm}[ht]
	\KwData{List of PUs, workload}
	\KwResult{Distribute the workload across these PUs and process it simultaneously}
	initialization\;\label{alg:scheduler_initialization}
	create as many threads (T) as PUs\;
	\ForEach{$T$}{%
		\While{not reached the end of data}{
			determine start and finish positions\;\label{alg:scheduler_chunk_determination}
			\uIf{pu.type is GPU}{\label{alg:scheduler_pu_type_GPU}
				\tcc{explicit data management is performed using CUDA API calls}
				create device variables\;\label{alg:scheduler_create_device_variables}
				allocate device memory\;\label{alg:scheduler_gpu_start}
				copy chunks of data from host to device\;
				execute CUDA kernel \tcp*{Example: Listing \ref{lst:GPU_triad}}
				copy data back to host memory\;
				free memory \;\label{alg:scheduler_gpu_finish}
			}
			\uElseIf{pu.type is MIC}{ \label{alg:scheduler_pu_type_MIC}
				\tcc{explicit data management is done using LEO \textit{in} and \textit{out} clauses}
				offload data and control to MIC device \tcp*{Example: Listing \ref{lst:MIC_triad}}
			}
			\uElseIf{pu.type is CPU}{\label{alg:scheduler_pu_type_CPU}
				\tcc{no explicit data management is needed}
				execute the CPU code \tcp*{Example Listing \ref{lst:CPU_triad}}
			}
		}
	}
	\caption{The algorithm used for runtime scheduling and workload distribution.}
	\label{alg:scheduler}
\end{algorithm}

\section{Evaluation}
\label{sec:evaluation}

In this section, we first describe the experimentation environment, including the hardware configuration, application benchmarks, and the data-set used for evaluation of our approach. Thereafter, we discuss the results of the study.

\subsection{Experimentation environment}
\label{sec:exp_environment}

For evaluation of our approach, we have used applications from the industry standard STREAM and STREAM2 benchmarks \cite{McCalpin1995}. We vary the stream size, chunk size, and the number of available resources. Details about the system configuration, application benchmark, considered data-sets, and the measurement metrics will follow.

\subsubsection{System configuration}
\label{sec:sys_config}

We used our heterogeneous system named \textit{DISA}, which comprises two Intel Xeon Gold CPUs, and four Quadro P4000 GPUs. Please note that in our experiments the CPU hyper-threading is disabled. Table \ref{tab:sys_config} lists the details of our systems. For experimental evaluation, we vary the number of resources used in the DISA system, such as CPU only, 1GPU, 2GPUs, 3GPUs, 4GPUs, CPU+1GPU, CPU+2GPUs, CPU+3GPUs, and CPU+4GPUs.

\begin{table}[ht]
	\centering
	\scriptsize
	\caption{The system configuration details for \textit{DISA}.}
	\label{tab:sys_config}
	\begin{tabular}{@{}p{2.5cm}p{2.8cm}p{2.8cm}@{}}
		\toprule
		Specs			& Intel Xeon	& NVIDIA GPU	\\ \midrule
		Type			& Gold 6148		& Quadro P4000	\\
		Frequency (GHz)	& 2.4 - 3.7		& 1.48			\\
		\# of Cores		& 20            & 1792 			\\
		\# of Threads	& 40            & /          	\\
		Cache (MB)		& 27.5			& /          	\\
		Memory (GB)		& 768   		& 8      		\\
		TDP (W)			& 150     		& 105      		\\ \bottomrule
	\end{tabular}
\end{table}

\subsubsection{Benchmark application and data-set}
\label{sec:benchmark}

We use applications from the STREAM and STREAM2 benchmarks to evaluate our approach. We use the HSTREAM source-to-source compiler to generate the heterogeneous version of the STREAM benchmark \cite{McCalpin1995}. We use the \textit{COPY, SCALE, SUM}, and \textit{TRIAD} functions from the STREAM benchmark, and the \textit{FILL} and \textit{DAXPY} functions from the STREAM2 benchmark. Details and information about the considered functions of the STREAM benchmark are available on-line \footnote{http://www.cs.virginia.edu/stream/}.

The stream array size is varied between 256, 512, 1024, 2048, 4096, and 8192MB to simulate various scenarios of the workload. Furthermore, when we split the workload among the available PUs, we vary the chunk sizes between 1, 2, 4, 8, 16, 32, and 64MB. The chunk size indicates the amount of data that should be sent for processing to a specific PU.

To address the variability of the results, we repeat each experiment 10 times and report the average value.

\subsubsection{Metrics}
\label{sec:metrics}

In our experiments, we consider two aspects: (1) the programming productivity and (2) the performance.
We measure the programming productivity with respect to the total lines of code and the lines of code that are specific to a programming framework required to parallelize the code. We use our tool, named CodeStat \cite{memeti2017benchmarking} to measure the programming productivity.
With respect to the performance, we measure the throughput, which reflects the amount of data that can be processed within a time unit.

\subsection{Results}
\label{sec:results}

In this section, we first describe the evaluation results with respect to programming productivity. Thereafter, we describe the results with respect to performance.

\subsubsection{Programming productivity}

Table \ref{tab:loc} shows the programming productivity expressed in lines of code (LOC). We consider the total LOC, and LOC specific to a programming language, such as OpenMP, and CUDA. For HSTREAM input we also show the LOC required to create the platform description file.
We compare different versions of implementations of the STREAM benchmark, including sequential, multi-core version using OpenMP, accelerated version using CUDA, and the heterogeneous version using HSTREAM. The last row shows the LOC of the HSTREAM generated version.

We may observe that the multi-core version of STREAM benchmark requires 8 OpenMP specific lines of code to be added, whereas the accelerated version requires about 55 CUDA specific lines of code. To parallelize the STREAM benchmark with HSTREAM language extension there are needed only 8 HSTREAM specific LOC, exactly the same as with OpenMP. For HSTREAM the developer needs to provide the PDL file as well, the LOC for which depends on the system. For the DISA system, the PDL file has 57 lines of code. The generated code, which is what a programmer would need to write manually for execution on heterogeneous systems that comprise host CPUs, and accelerating devices such as GPUs and Intel Xeon Phis, requires in total 1195 LOC, of which 69 are OpenMP and 131 are CUDA specific lines of code.

We may conclude that HSTREAM maintains the same level of programming complexity as OpenMP, at the cost of providing an XML-based description of the platform.

\begin{table}[]
	\centering
	\caption{The programming effort expressed in lines of code (LOC) required to program the STREAM benchmark using different programming frameworks.}
	\label{tab:loc}
	\begin{tabular}{@{}p{3.5cm}p{0.9cm}p{0.9cm}p{0.9cm}p{0.9cm}@{}}
		\toprule
						  			& Total 	  & OpenMP 	   & CUDA     & Other     \\ \midrule
		sequential (C)        		& 131       & 0          & 0        & 0         \\
		multi-core (OpenMP)     	& 214       & 8          & 0        & 0         \\
		accelerator (CUDA) 			& 190       & 0          & 55       & 0         \\
		multi-core + acc (HSTREAM) 	& 210       & 8          & 0        & 57        \\ \midrule
		HSTREAM generated 			& 1195      & 69         & 131      & 0         \\ \bottomrule
	\end{tabular}
\end{table}

\subsubsection{Performance}

\begin{figure}[ht]
	\centering
	\includegraphics[width=\linewidth]{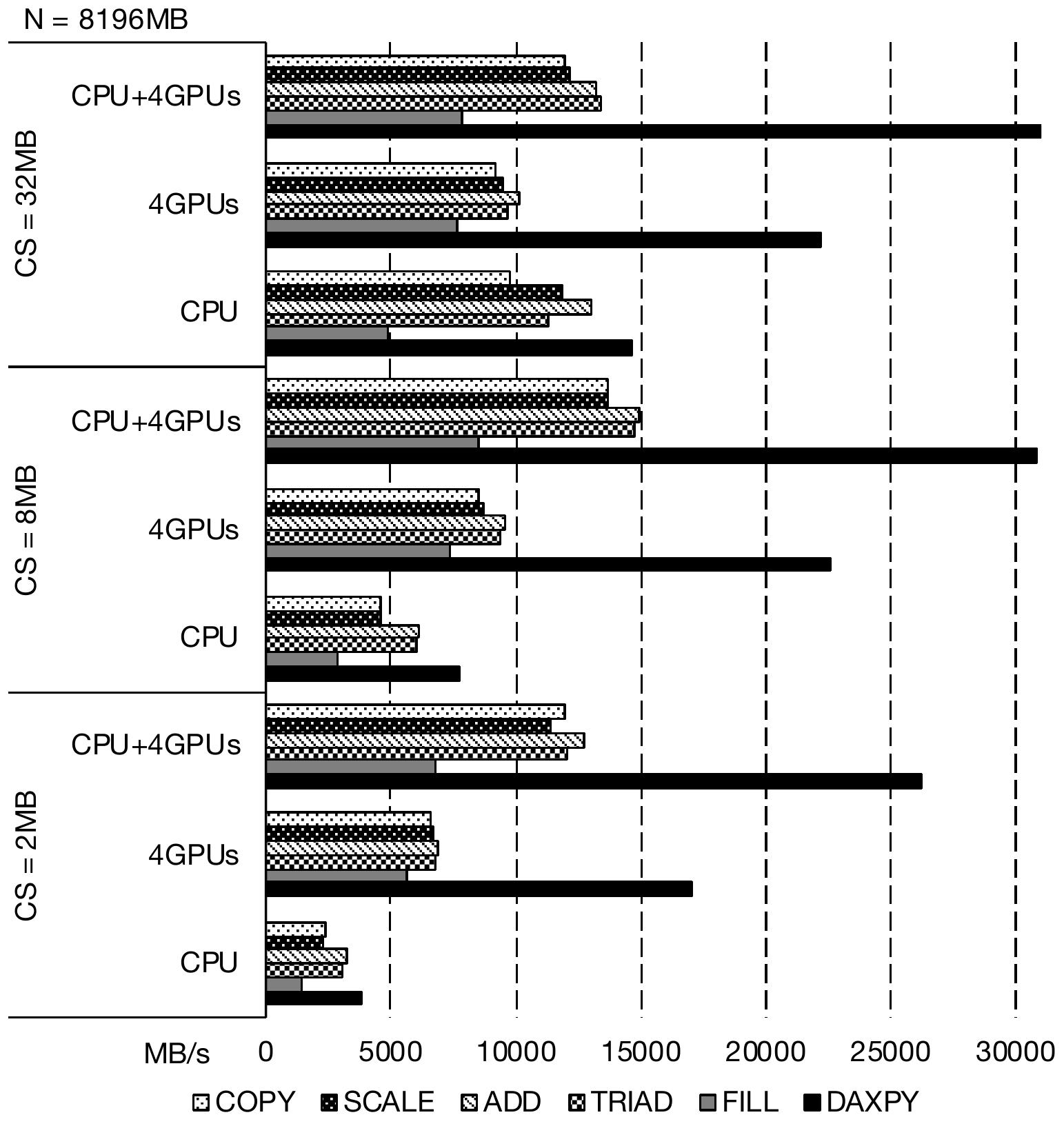}
	\caption{The performance of heterogeneous STREAM benchmark when using host CPUs only, 4GPUs, and CPU+4GPUs on our DISA system. We vary the chunk size between 2, 8, and 32MB. The input size used in this experiment is 8192MB.}
	\label{fig:performance}
\end{figure}

Figure \ref{fig:performance} depicts the throughput (MB/s) of the selected functions from the STREAM and STREAM2 benchmark, including \textit{COPY, SCALE, ADD, TRIAD, FILL,} and \textit{DAXPY}. We vary the stream size between 256, 512, 1024, 2048, 4096, and 8192MB. From our experiments, we have observed that the stream size does not impact the throughput, therefore we show the results only for the largest stream size (that is 8192MB). We also vary the chunk size (CS) between 1, 2, 4, 8, 16, 32, and 64MB. We vary the number of processing units engaged for computation, such as CPU only, 1GPU, 2GPUs, 3GPUs, 4GPUs, CPU+1GPU, CPU+2GPUs, CPU+3GPUs, and CPU+4GPUs. However, due to space limitation, we only show results for CS 2, 8, and 32MB, and system configurations CPU, 4GPUs, and CPU+4GPUs.

We may observe that in all cases, the execution that uses all available resources (CPU+4GPUs) results with the highest throughput. With respect to the chunk size, we may observe that the throughput depends on the type of processing units engaged in the execution, and the type of computations. For example, the more complex functions such as TRIAD and DAXPY benefit more when executed in accelerated devices, whereas simpler functions such as FILL and ADD perform better on CPUs. The highest throughput observed in our experiments is when executing the DAXPY function on CPU+4GPUs with a chunk size of 32MB.

\section{Related work}
\label{sec:rw}

Recent versions of OpenMP and OpenACC support offloading computations to accelerating devices \cite{xu2013}. Offloading to a single accelerating device can be easily achieved by using simple compiler-directives, whereas offloading to multiple devices requires additional programming effort. 
HSTREAM enables developers to use high-level compiler directives (similar to OpenMP and OpenACC) to develop applications that can be executed on multiple heterogeneous devices, such as multi-core CPUs, NVIDIA GPUs, and Intel Xeon Phi accelerators.

\citet{zhang2011gstream} present GStream, a framework for data-streaming on systems accelerated with GPUs. GStream provides an application interface that software developers can use to express the parallelism of their streaming application without explicitly writing MPI messaging or CUDA memory copy instructions. Similar to our approach, the aim is to reduce the development time. Rather than exposing system developers to a new application programming interface, we have decided to introduce a new simple OpenMP like directive that developers can use to parallelize their streaming applications.

\citet{yan2017homp} propose a language extension to OpenMP for data-parallel processing. Their approach distributes data and computations of parallel loops across multiple PUs that may be homogeneous or heterogeneous. In comparison, we describe an OpenMP language extension for stream computing.

\citet{pop2011stream} propose a stream computing extension to the OpenMP programming model. Their extension decomposes programs into tasks and in an explicit manner provides instructions on how data should flow among these tasks. In comparison to their work, which targets multi-core architectures, we provide stream computing support for heterogeneous parallel computing systems that may comprise multiple multi-core processors and many-core accelerating devices.

\citet{del2017generic} and \citet{ernstsson2018} propose pattern based high-level application programming interface for stream and data computing on heterogeneous parallel computing systems. While template and skeleton based libraries may be helpful for generic applications, they are not recommended in case programmers want to use data-structures or algorithms optimized for a particular type of problem.

\citet{zhang2018auto} propose an auto-tuning approach for stream applications running on systems accelerated with Intel Xeon Phi. The authors exploit the pipeline parallelism through temporal sharing, which means that computations are overlapped with communication (data transfer from host to device, and vice-versa). We employ similar techniques in our solution, but in comparison we support heterogeneous systems accelerated with GPUs as well.

With the aim to alleviate the programming of heterogeneous systems, several source-to-source compilers \cite{beach2009,fonseca2013,ansel2009petabricks,rossbach2013} are proposed that can generate target specific code from a high-level representation. In comparison, our approach targets streaming applications. Furthermore, rather than introducing a new programming language, our solution extends a well-established programming model, such as OpenMP, to enable acceleration of streaming applications.

\section{Conclusion and future work}
\label{sec:conclusion_fw}

We presented our HSTREAM language extension to support stream computing on heterogeneous parallel computing systems. HSTREAM source-to-source compiler can automatically generate device-specific code, such as OpenMP for CPUs, CUDA for GPUs, and Intel Language Extension for Offloading for Intel Xeon Phi, from a high-level source code annotated with OpenMP-like compiler directives. The HSTREAM runtime is responsible to distribute the workload across the engaged processing units accordingly. We have evaluated the usefulness of our HSTREAM solution for stream computing in heterogeneous parallel computing systems with the STREAM benchmark. We have observed that while HSTREAM keeps the same programming simplicity as OpenMP, the generated code outperforms the CPU and GPU only versions of the code.

Future work may focus on extending the HSTREAM runtime to support dynamic and adaptive workload scheduling. Furthermore, we aim to extend the source-to-source compiler to support additional accelerating devices (such as FPGAs) and programming frameworks (such as OpenCL).


\end{document}